# Two are Better Than One: Adaptive Sparse System Identification using Affine Combination of Two Sparse Adaptive Filters

Guan Gui, Shinya Kumagai, Abolfazl Mehbodniya, and Fumiyuki Adachi
Department of Communications Engineering
Graduate School of Engineering
Tohoku University, Sendai, Japan
{gui, kumagai, method}@mobile.ecei.tohoku.ac.jp; adachi@ecei.tohoku.ac.jp

*Abstract*—**Sparse system identification problems often exist in many applications, such as echo interference cancellation, sparse channel estimation, and adaptive beamforming. One of popular adaptive sparse system identification (ASSI) methods is adopting only one sparse least mean square (LMS) filter. However, the adoption of only one sparse LMS filter cannot simultaneously achieve fast convergence speed and small steady-state mean state deviation (MSD). Unlike the conventional method, we propose an improved ASSI method using affine combination of two sparse LMS filters to simultaneously achieving fast convergence and low steady-state MSD. First, problem formulation and standard affine combination of LMS filters are introduced. Then an approximate optimum affine combiner is adopted for the proposed filter according to stochastic gradient search method. Later, to verify the proposed filter for ASSI, computer simulations are provided to confirm effectiveness of the proposed filter which can achieve better estimation performance than the conventional one and standard affine combination of LMS filters.**

*Keywords—least mean square, affine combination, sparse adaptive filter, adaptive system identification, $\ell_0$-norm sparse function.*

## I. INTRODUCTION

### A. Background and motivation

Adaptive system identification (ASI) includes many applications such as echo interference cancellation, sparse channel estimation, and adaptive beamforming. One of classical algorithms is called least mean square (LMS) which was first proposed by Widrow and Hoff [1]. The LMS filter is widely used in many applications which require tradeoff between convergence speed and steady-state mean square error (MSE). In other words, a faster (slower) convergence speed of LMS filter often yields a higher (lower) steady-state MSE and/or a higher steady-state mean square deviation (MSD). Unfortunately, LMS filter unable trades off between them due to adopting only one fixed step-size.

To deal with this problem, combination structure of two standard LMS filters, as shown in Fig. 1, is attracting a lot of attention in the last decades. The first adaptive filter $\mathbf{w}_1$ uses larger step-size than second filter $\mathbf{w}_2$ so that the combination filter can achieve a good/fair tradeoff between convergence speed and steady-state MSE. An improved filter using convex combination of two fixed step-size based standard LMS (CC-LMS) filters was first proposed [2] and later, its steady-state performance was analyzed in [3]. Moreover, an affine combination of two standard LMS (AC-LMS) filters was also proposed and was studied via transient MSE in [4]. However, both CC-LMS and AC-LMS filters do not consider the sparse structure of finite impulse response (FIR) of unknown systems.

In many scenarios, FIR of unknown systems are modeled to be sparse as shown in Fig. 2, containing only a few large coefficients (active) interspersed among many negligible ones (inactive). Taking advantage of such sparse prior information can improve the identifying performance. However, the proposed two combination structure filters [2] [4] do not exploit such information due to the fact that they adopted standard LMS filters. Thus, there is a great interest in exploiting the sparse structure information to improve the filtering performance in sparse systems.

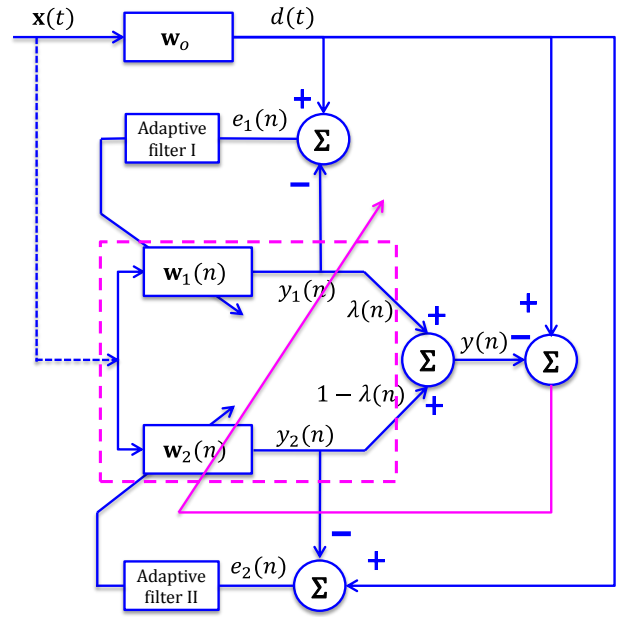

Fig. 1. Adaptive system identification utilizing an affine combination of two LMS filters to estimate the unknown system.

Motivated by the compressive sensing (CS) [5], [6], Chen and his collaborators proposed zero-attracting LMS filter (ZA-LMS) and reweighted ZA-LMS filter (RZA-LMS) using $\ell_1$- norm sparse penalty [7]. Based on this work, Taheri and Vorobyov proposed an improved sparse LMS filter using $\ell_p$-norm sparse penalty [8], which is termed as LP-LMS. Gu and his collaborators also proposed an improved sparse LMS filter using approximated $\ell_0$-norm sparse penalty [9], which is termed as L0-LMS. However, the above mentioned adaptive sparse LMS filters adopt only one filter.

To the best of our knowledge, no paper has reported the combined structure of two sparse LMS filters for ASSI. Based on the proposed affine combination filter in [4], in this paper, we propose a novel sparse combination filter adopting two L0-LMS filters to simultaneously achieve three merits: 1) fast convergence speed, 2) low steady-state MSE, and 3) exploiting system sparsity.

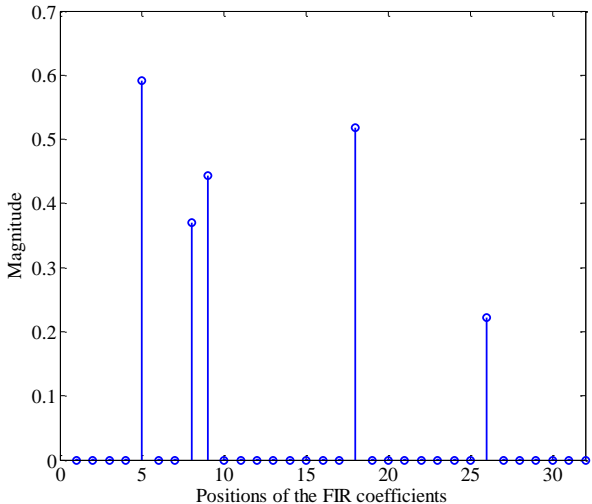

Fig. 2. A typical example of sparse system with FIR length 32 and 5 nonzero coefficients

### B. Main contributions

In this paper, we propose a novel sparse filter using affine combination of the two sparse LMS filters. The proposed filter has two properties: 1) Two LMS filters can provides a good tradeoff between convergence speed and steady-state MSD performance compared with traditional ASI method with only one LMS filter, 2) AC-L0LMS can exploit system sparsity information and then it can further improve the identification performance compared with AC-LMS filter.

The main contribution of this paper is summarized as follows: $\ell_0$-norm sparse function is introduced to cost function of standard AC-LMS filter and then, sparse AC-LMS filter (i.e., AC-L0LMS filter) is proposed for ASSI. Later, several representative experiments are conducted to confirm the effectiveness of our propose methods. In the first experiment, the steady-state MSD performance of the proposed affine combination filter is evaluated with the number of dominant FIR coefficients as parameter. In the second experiment, assuming the constant number of dominant FIR coefficients, the MSD performance of propose algorithms is evaluated with the step-size ratio as a parameter.

### C. Organizations and notations

The reminder of the rest paper is organized as follows. Section 2 reviews the affine combination of two standard LMS filters and problem formulation. In Section 3, we propose affine combination of two sparse LMS filters to improve system identification performance without sacrificing convergence speed. In section 4, the simulation results via MSD metric are presented to confirm the effectiveness of proposed filters, using a Monte-Carlo technique. Concluding remarks are presented in Section V.

Throughout the paper, matrices and vectors are represented by boldface upper case letters and boldface lower case letters, respectively; the superscripts $(\cdot)^T$, $(\cdot)^H$, and $(\cdot)^{-1}$ denote the transpose, the Hermitian transpose, and the inverse operators, respectively; E$\{\cdot\}$ denotes the expectation operator; $\|\mathbf{w}\|_0$ is the $\ell_0$-norm operator that counts the number of nonzero taps in $\mathbf{w}$; $\mathbf{w}_o$ and $\mathbf{w}(n)/\widetilde{\mathbf{w}}(n)$ are the actual system FIR and its $n$-th iterative adaptive FIR estimator, respectively. E$\{\cdot\}$ denotes the expectation operator.

## II. SYSTEM MOLDE AND PROBLEM FORMULATION

Consider a typical sparse finite impulse response (FIR) system, as shown in Fig. 2, the input signal $\mathbf{x}(t)$ and ideal output signal $d(t)$ are related by

$$d(t) = \mathbf{w}_o^T \mathbf{x}(t) + z(t), \tag{1}$$

where $\mathbf{w}_o = [w_1, w_2, \ldots, w_N]^T$ is a $N$-length unknown system FIR vector which is supported only by $K$ ($K \ll N$) dominant coefficients; $x(t) = [x(t), x(t-1), \ldots, x(t-N+1)]^T$ is $N$-length input signal vector and $z(t)$ is assumed zero-mean and independent and identically distributed (i.i.d.) random noise variable in the system. The object of ASI is to identify the unknown sparse FIR coefficients $\mathbf{w}_o$ using the input signal $\mathbf{x}(t)$ and output signal $y(t)$. According to Eq. (1), system identification error $e_i(n)$ is given by

$$e_i(n) = d(t) - \mathbf{w}_i^T(n)\mathbf{x}(t), i = 1,2, \tag{2}$$

where $\mathbf{w}_i(n)$ is the $n$-th adaptive FIR estimator and $d(n) = $. Based on Eq. (2), cost function of the $i$-th standard LMS filter [10] is given by

$$G_i(n) = \frac{1}{2} e_i^2(n), i = 1, 2. \tag{3}$$

With different step-size $\mu_i$, then the $(n + 1)$-th update FIR system vector is updated according to

$$\begin{aligned}\mathbf{w}_i(n+1) &= \mathbf{w}_i(n) + \mu_i \nabla_{\mathbf{w}_i(n)} G_i(n) \\ &= \mathbf{w}_i(n) + \mu_i e_i(n)\mathbf{x}(t), i = 1,2,\end{aligned} \tag{4}$$

where $\mathbf{w}_i(n), i = 1,2$ are the $N$-dimensional adaptive FIR vectors. Without loss of generality, we assume $\mu_1 \geq \mu_2$ so that $\mathbf{w}_1(n)$ achieves faster convergence speed than $\mathbf{w}_2(n)$. Note that the steady-state MSD performance of the $\mathbf{w}_2(n)$ is better than $\mathbf{w}_1(n)$. Also, assuming both $\mathbf{w}_1(n)$ and $\mathbf{w}_2(n)$ are coupled deterministically and statistically through input signal vector $\mathbf{x}(t)$ and additive noise variable $z(t)$. Since $y_i(n) = \mathbf{w}_i^T(n)\mathbf{x}(t), i = 1, 2$, denotes individual output signal from $i$-th LMS filter, according to Fig. 2, the system output signal $y(n)$ of the affine combination of the two LMS filters is given by

$$\begin{aligned}y(n) &= \lambda(n) y_1(n) + (1 - \lambda(n))y_2(n) \\ &= \lambda(n)\mathbf{w}_1^T(n)\mathbf{x}(n) + (1 - \lambda(n))\mathbf{w}_2^T(n)\mathbf{x}(t),\end{aligned}$$

$$= \{\lambda(n)[\mathbf{w}_1(n) - \mathbf{w}_2(n)] + \mathbf{w}_2(n)\}^T \mathbf{x}(t)$$
$$= \{\lambda(n)\mathbf{w}_{12}(n) + \mathbf{w}_2(n)\}^T \mathbf{x}(t), \quad (5)$$

where $\mathbf{w}_{12}(n) := \mathbf{w}_1(n) - \mathbf{w}_2(n)$ is a difference filter and $\lambda(n)$ is a affine combination parameter to decide final system identification error. In Eq. (5), we can find that $y(n)$ can be considered as a combination of filter ($\mathbf{w}_2(n)$) and a weighted filter ($\lambda(n)\mathbf{w}_{12}(n)$), the equivalent filter can be given by

$$\mathbf{w}_{eq}(n) := \lambda(n)\mathbf{w}_{12}(n) + \mathbf{w}_2(n). \quad (6)$$

According to (1) and (5), the overall system error is given by

$$e(n) = d(t) - y(n)$$
$$= [\mathbf{w}_{o2}(n) - \lambda\mathbf{w}_{12}(n)]^T \mathbf{x}(t) + z(t), \quad (7)$$

where $\mathbf{w}_{o2}(n) := \mathbf{w}_o - \mathbf{w}_2(n)$ is also considered as a difference filter. This setup generalizes the combination of adaptive filter outputs, and can be used to study the properties of the optimal combination. In [4], the authors proposed the optimal affine combiner

$$\lambda_o(n) = \frac{\mathbf{w}_{o2}^T(n)\mathbf{R}_{xx}\mathbf{w}_{12}(n)}{\mathbf{w}_{12}^T(n)\mathbf{R}_{xx}\mathbf{w}_{12}(n)}, \quad (8)$$

which is the expectation for the optimum $\lambda(n)$, as a function of the unknown weight vector $\mathbf{w}_o$, where $\mathbf{R}_{xx} = E\{\mathbf{x}(t)\mathbf{x}^T(t)|\mathbf{w}_2(n), \mathbf{w}_{12}(n)\}$ denotes the input conditional autocorrelation matrix. It is easy found that the optimal affine combiner is based on prior knowledge of the unknown system FIR $\mathbf{w}_o$. However, it cannot be utilized in reality. Employing a stochastic gradient search method, suboptimal affine combiner $\lambda_s(n)$ [4] was proposed as follows

$$\lambda_s(n+1) = \lambda_s(n) + \mu_\lambda[d(t) - \widehat{\mathbf{w}}_{12}^T(n)\mathbf{x}(t)]\mathbf{w}_{12}^T(n)\mathbf{x}(t), (9)$$

where $\widehat{\mathbf{w}}_{12} = \lambda_s(n)\mathbf{w}_1(n) + [1 - \lambda_s(n)]\mathbf{w}_2(n)$ and $\mu_\lambda$ are parameters for tracking the adaptation of $\mathbf{w}_1(n)$ and $\mathbf{w}_2(n)$.

III. AFFINE COMBINATION OF TWO SPARSE LMS FILTERS

Since the affine combination of two standard LMS filters neglects the inherent system sparsity, it often causes performance loss. Unlike the traditional method, we propose an affine combination of two L0LMS filters to exploit the system sparsity, with individual cost functions [9] given as by,

$$L_i(n) = \frac{1}{2}e_i^2(n) + \beta_i \|\mathbf{w}_i(n)\|_0, i = 1, 2. \quad (10)$$

It is well known that solving the $\|\mathbf{w}_i(n)\|_0$ in (10) is a (non-deterministic polynomial-time) NP-hard problem [5]. To deal with this problem, we approximate it by a continuous function

$$\|\widetilde{\mathbf{w}}_i(n)\|_0 \approx \sum_{k=0}^{N-1}(1 - e^{-\alpha|w_{i,k}(n)|}), i = 1,2. \quad (11)$$

According to (11), cost function of the $i$-th L0LMS filter can be changed to

$$L_i(n) = \frac{1}{2}e_i^2(n) + \beta_i \sum_{k=0}^{N-1}(1 - e^{-\alpha|w_{i,k}(n)|}), i = 1,2. \quad (12)$$

Similarly, with different step-size $\mu_i$, the $(n+1)$-th update sparse FIR vector is derived as

$$\widetilde{\mathbf{w}}_i(n+1) = \widetilde{\mathbf{w}}_i(n) + \mu_i \nabla_{\mathbf{w}_i(n)} L_i(n)$$
$$= \widetilde{\mathbf{w}}_i(n) + \mu_i e_i(n)\mathbf{x}(t)$$
$$+ \alpha\mu_i\beta_i \text{sgn}\{\widetilde{\mathbf{w}}_i(n)\}e^{-\alpha|w_i(n)|}, \quad (13)$$

for $i = 1,2$. It is worth mentioning that the exponential function in Eq. (13) will cause high computational complexity. To reduce the high complexity, the first order Taylor series expansion of exponential function is taken into consideration as

$$e^{-\alpha|\widetilde{w}_i(n)|} \approx \begin{cases} 1 - \alpha|w_i(n)|, \text{when } |\widetilde{w}_i(n)| \leq 1/\alpha \\ 0, \quad \text{others} \end{cases}. \quad (14)$$

It is worth mentioning that the positive parameter $\alpha$ controls the system sparseness and identification performance. Though the L0LMS can exploit system sparsity on adaptive system identification, unsuitable parameter $\alpha$ will cause overall identification performance degradation. In this paper, we adopted $\alpha = 10$ which is also suggested as in [11]. According to above analysis, the modified update equation of L0LMS can be rewritten as

$$\widetilde{\mathbf{w}}_i(n+1) = \widetilde{\mathbf{w}}_i(n) + \mu_i e_i(n)\mathbf{x}(t) - \alpha\mu_i S\{\widetilde{\mathbf{w}}_i(n)\}, \quad (15)$$

where the $\ell_0$-norm sparse penalty approximation function $S\{\widetilde{\mathbf{w}}_i(n)\}$ is defined as

$$S\{\widetilde{\mathbf{w}}_i(n)\} = \begin{cases} 2\alpha^2\widetilde{w}_i(n) - 2\alpha\text{sgn}\{\widetilde{w}_i(n)\}, \text{if } |\widetilde{w}_i(n)| \leq \frac{1}{\alpha} \\ 0, \quad \text{others}. \end{cases}$$

Analogy to the Eq. (9), suboptimal affine combiner $\widetilde{\lambda}_s(n)$ can also be given by

$$\widetilde{\lambda}_s(n+1) = \widetilde{\lambda}_s(n) + \widetilde{\mu}_\lambda[d(t) - \widehat{\widetilde{\mathbf{w}}}_{12}^T(n)\mathbf{x}(t)]\widetilde{\mathbf{w}}_{12}^T(n)\mathbf{x}(t), \quad (16)$$

where $\widehat{\widetilde{\mathbf{w}}}_{12} = \widetilde{\lambda}_s(n)\widetilde{\mathbf{w}}_1(n) + [1 - \widetilde{\lambda}_s(n)]\widetilde{\mathbf{w}}_2(n)$ and $\widetilde{\mu}_\lambda$ are parameters for tracking the adaptation of $\widetilde{\mathbf{w}}_1(n)$ and $\widetilde{\mathbf{w}}_2(n)$. By exploiting the system sparsity, output signal $\widetilde{y}(n)$ of affine combination of the two L0LMS filters is given by

$$\widetilde{y}(n) = \widetilde{\lambda}(n)\widetilde{y}_1(n) + \left(1 - \widetilde{\lambda}(n)\right)\widetilde{y}_2(n)$$
$$= \{\widetilde{\lambda}(n)\widetilde{\mathbf{w}}_{12}(n) + \widetilde{\mathbf{w}}_2(n)\}^T \mathbf{x}(t). \quad (17)$$

Finally, the overall system error is given by

$$\widetilde{e}(n) = d(t) - \widetilde{y}(n)$$
$$= [\widetilde{\mathbf{w}}_{o2}(n) - \lambda\widetilde{\mathbf{w}}_{12}(n)]^T \mathbf{x}(n) + z(t), \quad (18)$$

where $\widetilde{\mathbf{w}}_{o2}(n) := \mathbf{w}_o - \widetilde{\mathbf{w}}_2(n)$ is the difference filter between real and sparse FIR vector.

IV. COMPUTER SIMULATIONS

To confirm effectiveness of the proposed ASSI method using affine combination of two sparse NLMS filters, we evaluate their steady-state MSD performance which is defined by

$$\text{MSD}\{\widehat{\mathbf{w}}(n)\} = E\{\|\mathbf{w} - \widehat{\mathbf{w}}(n)\|_2^2\}. \quad (19)$$

The results are averaged over 1000 independent Monte-Carlo runs. The signal-to-noise ratio (SNR) is defined as $E_s/\sigma_n^2$, where $E_s$ is the received power of the input signal. Computer simulation parameters are listed in Table. I.

TAB. I. SIMULATION PARAMETERS.

| parameters | values |
|---|---|
| length of FIR length | $N = 32$ |
| no. of nonzero coefficients | $K = 3$ and $6$ |
| positions of nonzero coefficients | Random allocation |
| distribution of FIR coefficient | random Gaussian $\mathcal{CN}(0,1)$ |
| SNR | $E_s/\sigma_n^2 = 10\text{dB}$ |
| step-size of filter I and filter II | $\mu_1 = \frac{1}{N+\gamma}$ and $\mu_2 = \delta\mu_1$ |
| controlling the ratio $\delta = \mu_1/\mu_2$ | $\delta \in (0,1]$ |
| controlling $\mu_1$ of the filter I | $\gamma = 4$ |
| tracking adaptation of $\mathbf{w}_1$ and $\mathbf{w}_2$ in $\lambda_s(n)$ | $\mu_\lambda = 1$ |
| parameters for $\ell_0$-norm sparse penalty | $\beta = 0.02\sigma_n^2$ and $\alpha = 10$ |

In the first example, considering two different step-size ratios $\delta \in \{0.3, 0.5\}$, steady-state MSD performance of proposed filter is evaluated for $K = 3$ and $6$, respectively, in Figs. 3-6. To verify the effectiveness of the proposed ASSI method, we compare it with three previous methods, i.e., LMS [10], L0LMS [9] and AC-LMS [4]. For a fair evaluation of these algorithms, same regularization parameter is adopted for L0LMS and AC-L0LMS algorithms, i.e., $\beta = 0.02\sigma_n^2$, which is also recommended by [12][13]. As Figs. 3-6 show, our proposed filter can achieve lower MSD performance without reducing the convergence speed. Note that choosing smaller (bigger) $\delta$ can achieve lower (higher) MSD and (faster) slower convergence speed. In addition, MSD performance of the proposed filter also depends on FIR sparseness in real systems. For sparser system, the proposed filter can achieve much lower MSD performance by comparing MSD curves of proposed filter in Fig. 3 ($K = 3$) and Fig. 4 ($K = 6$). Hence, the proposed filter can choose different parameters (e.g., $\delta$ and $\beta$) to meet concrete requirements of the system.

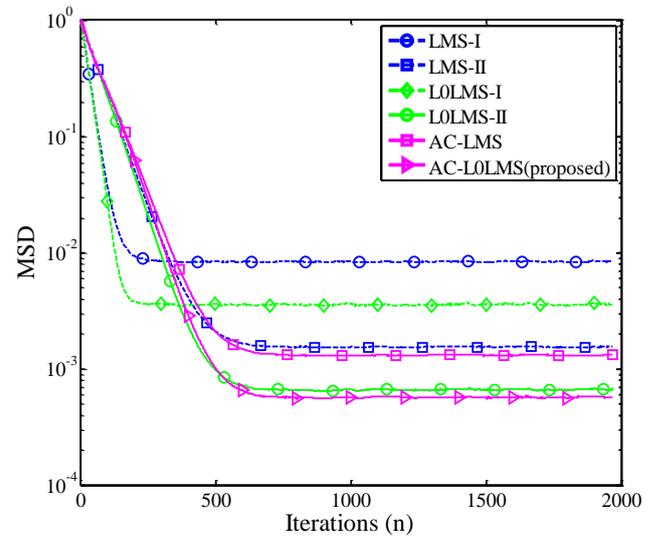

Fig. 4. MSD versus iterations at $K = 6$, $\gamma = 4$ and $\delta = 0.3$ (SNR=10dB).

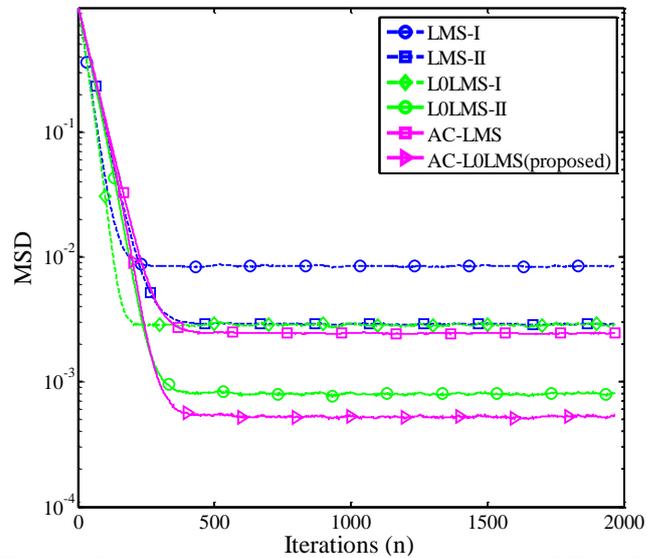

Fig. 5. MSD versus iterations at $K = 3$, $\gamma = 4$ and $\delta = 0.5$ (SNR=10dB).

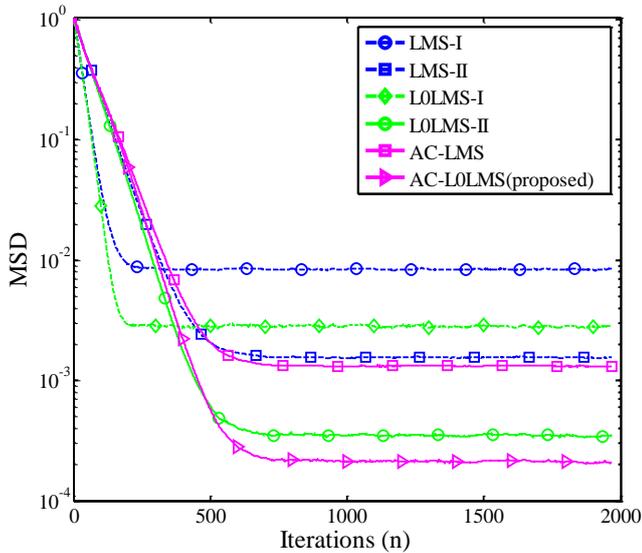

Fig. 3. MSD versus iterations at $K = 3$, $\gamma = 4$ and $\delta = 0.3$ (SNR=10dB).

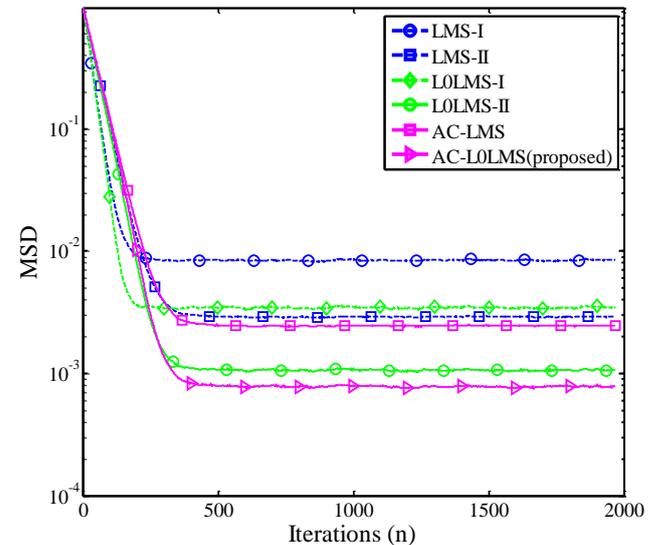

Fig. 6. MSD versus iterations at $K = 6$, $\gamma = 4$ and $\delta = 0.5$ (SNR=10dB).

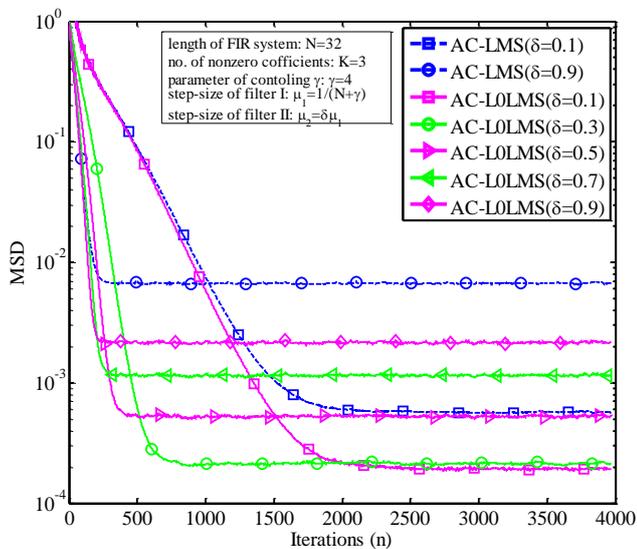

Fig. 7. MSD performance evaluations versus different $\delta$ at $K = 3$.

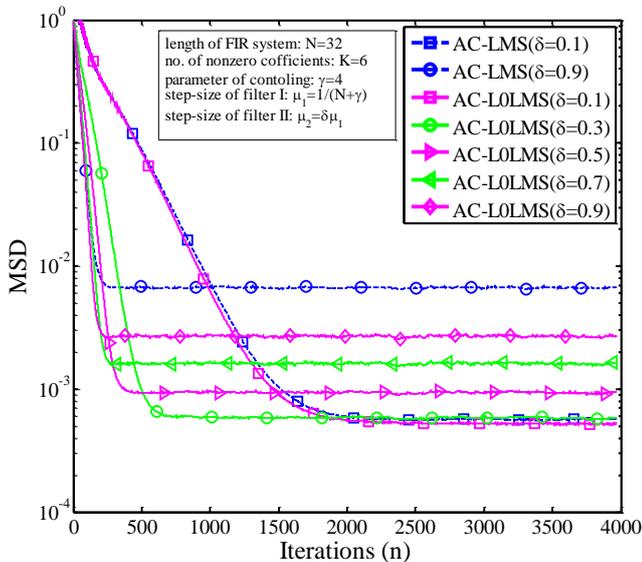

Fig. 8. MSD performance evaluations versus different $\delta$ at $K = 6$.

In the second example, system performance using proposed filter is also evaluated with respect to different $\delta$ ratio which controls the filter II. Note that the step-size of filter I was fixed $\mu_1 = 1/(N + \gamma)$ and the step-size of filter II was set as $\mu_2 = \delta\mu_1$, where $\delta \in (0,1]$. It is well known that LMS filter using large (small) step-size obtains low (high) steady-state MSE performance with fast (slow) convergence speed. The length of FIR system, $\gamma$ and $\delta$ are critical parameters which control the convergence speed and steady-state MSE performance of the proposed filter. Five ratios $\delta \in \{0.1, 0.3, 0.5, 0.7, 0.9\}$ are adopted as shown in Figs. 7 and 8. In the two figures, $\delta = 0.1$ and $\delta = 0.9$ are utilized in standard AC-LMS filter as performance benchmarks. We observe that the proposed filter can achieve better a performance than either standard AC-LMS filter or single LMS filter.

## V. CONCLSION

Traditional ASSI methods often apply only one sparse LMS filter with an invariant step-size which cannot balance well between steady-state MSD performance and convergence speed. Hence, they are vulnerable to either performance loss or convergence speed deceleration. In other words, they cannot simultaneously achieve fast convergence speed and high steady-state MSD performance. Unlike these traditional methods, in this paper, we proposed an affine combination of two sparse LMS filters which can achieve fast convergence and high steady-state MSD performance to improve ASSI performance. First, problem formulation and standard affine combination of LMS filters were introduced. Then, $\ell_0$-norm sparse constraint function based affine combination of two sparse LMS filters for ASSI was presented. System identification performance depends on which affine combiner to choose. The approximate optimum affine combiner was adopted for the proposed filter according to stochastic gradient search method. Later, to verify the effectiveness of the proposed filter for ASSI, selected simulations were provided to confirm the effectiveness of the proposed filter which can achieve better estimation performance than the conventional one and standard affine combination of LMS filters.

ACKNOWLEDGMENT

This work was supported by grant-in-aid for the Japan Society for the Promotion of Science (JSPS) fellows grant number 24·02366.